\documentclass{ifacconf}

\usepackage[round]{natbib} 
\begin{document}
\begin{frontmatter}

\title{Bases for spin systems and qudits from angular momentum theory}

\author[First]{Maurice R. Kibler}$^{1, 2}$

\address[First]{$^1$Universit\'e de Lyon, F-69622, Lyon, France $^2$Universit\'e
Lyon 1, Villeurbanne, CNRS/IN2P3, UMR5822, Institut de Physique Nucl\'eaire 
de Lyon (e-mail: kibler@ipnl.in2p3.fr)}

\begin{abstract}
Spin bases of relevance for quantum systems with cyclic symmetry 
as well as for quantum information and quantum computation are 
constructed from the theory of angular momentum. This approach is 
connected to the use of generalized Pauli matrices (in dimension 
$d$) arising from a polar decomposition of the group SU$_2$. Numerous 
examples are given for $d=2, 3$ and $4$.
\end{abstract}


\end{frontmatter}

\section{INTRODUCTION}

Hilbert spaces of finite dimension $d$ play an important role for the quantum
mechanical description of dynamical systems and systems of qudits (qubits 
correspond to $d=2$, qudits to $d$ arbitrary). Generally speaking, bases for 
finite-dimensional subspaces of the representation space of the group SO$_3$ 
or its universal covering SU$_2$ can be used for spin systems (used in 
spectroscopy and quantum chemistry) and qudits (used in quantum information 
and quantum computation). Such bases can be constructed from tools developed 
for a study of supersymmetric quantum mechanics [1--4] and for a nonstandard
approach to the representations of SU$_2$ [5--7]. 

It is the object of this talk to describe a method for constructing 
bases which are of central importance in quantum information theory, 
namely, mutually unbiased bases (MUBs). In
dimension $d$, two bases are said to be unbiased if and only if the modulus 
of the inner product of any vector of one basis with any vector of the 
other one is equal $1/{\sqrt{d}}$ [8--17]. These bases are of paramount 
importance for quantum cryptography and quantum state tomography. 

We develop here an approach that gives a 
complete solution for the construction of MUBs in the case where the dimension 
$d$ of the considered Hilbert space is a prime number. This approach is based 
on a nonstandard approach to the theory of angular momentum viewed through the 
prism of the Lie algebra of SU$_2$. The concepts of Weyl pairs and generalized 
Pauli group useful in quantum computation and quantum information are also 
briefly discussed. 

Most of the material presented in this work takes 
its origin in papers published by the author 
and its collaborators [18--26].

The notation adopted here is the one used in 
quantum mechanics. Let us simply mention that 
$\delta _{a,b}$ stands for the Kronecker symbol for $a$ and 
$b$, $A^{\dagger}$ denotes the adjoint of the operator $A$, 
and $[A,B]$ is the commutator of the operators $A$ and $B$.

\section{SPIN BASES}

\subsection{Angular momentum states}
Let us consider a generalized angular momentum. We denote as $j^2$ and 
$j_z$ its square and $z$-component, respectively. The common orthonormalized 
eigenvectors of the two commuting operators $j^2$ and $j_z$ are written 
$| j , m \rangle$. We know that

\begin{eqnarray}
     j^2 |j , m \rangle = j(j+1) |j , m \rangle, 
\quad j_z |j , m \rangle = m      |j , m \rangle, 
\nonumber 
\end{eqnarray}

where $m = j, j-1, \ldots, -j$ and $2j \in {\bf N}$. For a fixed 
value of the quantum number $j$, we note ${\cal E}(2j+1) \sim {\bf C}^{2j+1}$ 
the $(2j+1)$-dimensional Hilbert space spanned by the basis 

\begin{eqnarray}
b_s = \{ |j , m \rangle : m = j, j-1, \ldots, -j \}. 
\nonumber 
\end{eqnarray}

The basis $b_s$ is 
adapted to spherical symmetry (adapted to the group SO$_3$ if $j \in {\bf N}$ 
or SU$_2$ if $2j \in {\bf N}$).  

In the applications to spectroscopy, the generalized angular momentum can be an 
orbital angular momentum, a spin angular momentum, a total (spin $+$ orbital) 
angular momentum, a nucleus spin, etc. The vectors $|j , m \rangle$ can thus 
have several realizations. For example, in the spectroscopy of a partly filled 
shell ion with configuration $n\ell^N$, we have state vectors of type 
$|J , M \rangle \equiv |n\ell^N \tau S L J M \rangle$ in the Russell-Saunders 
coupling (here $j = J$ and $m = M$).  

The vectors $|j , m \rangle$ can also serve for quantum information 
and quantum computation. By introducing the notation 
          
	  \begin{eqnarray}
k = j - m, \quad | k \rangle = | j , m \rangle, 
           \quad d = 2j+1, 
          \nonumber 
	  \end{eqnarray}  
	  
we get the vectors  $| 0   \rangle$   (for $m = j$), 
                    $| 1   \rangle$   (for $m = j-1$), $\ldots$, 
		    $| d-1 \rangle$   (for $m = -j$). 
The vectors  $| 0   \rangle$,   
             $| 1   \rangle$, $\ldots$, 
	     $| d-1 \rangle$ are called qudits.	Then, the basis $b_s$ becomes 

\begin{eqnarray}
B_d = \{ | k \rangle : k = 0, 1, \ldots, d-1 \},
\nonumber 
\end{eqnarray} 

which is referred to as the computational basis in quantum information 
theory. In $d$ dimensions, the most general qudit is written as 

\begin{eqnarray}
c_0 | 0 \rangle +  c_1 | 1 \rangle + \ldots + c_{d-1} | d-1 \rangle,
\nonumber 
\end{eqnarray} 

where $c_k \in {\bf C}$ for $k = 0, 1, \ldots, d-1$. The case 
$d = 2 \Leftrightarrow j = 1/2$ corresponds to qubits generated by  
$| 0 \rangle = | \frac{1}{2} ,   \frac{1}{2} \rangle$ and 
$| 1 \rangle = | \frac{1}{2} , - \frac{1}{2} \rangle$. Note that the vectors 
$| \frac{1}{2} ,   \frac{1}{2} \rangle$ and 
$| \frac{1}{2} , - \frac{1}{2} \rangle$ are nothing but the spinorbitals 
$\alpha$ (for spin up) and $\beta$ (for spin down), respectively, used in 
quantum chemistry.  

In the rest of this paper, we shall use both the notation $| k \rangle$ familiar 
in (i) quantum information and quantum computation and (ii) the description of
cyclic systems (for which $| d \rangle \equiv | 0 \rangle$, 
$| d+1 \rangle \equiv | 1 \rangle$, \ldots) and equally well the notation 
$| j , m \rangle$ employed in (i) angular momentum 
theory (from a physical point of view or in its approach from the group SU$_2$),
(ii) nuclear, atomic and molecular spectroscopy, and (iii) quantum 
chemistry. For $d = 2$, we shall play with the notation 

\begin{eqnarray}
\alpha = | \frac{1}{2} ,   \frac{1}{2} \rangle = | 0 \rangle
\nonumber 
\end{eqnarray} 

and 

\begin{eqnarray}
\beta  = | \frac{1}{2} , - \frac{1}{2} \rangle = | 1 \rangle.
\nonumber 
\end{eqnarray} 

\subsection{A noncanonical basis for SU$_2$}

We now define the operator $v_{0a}$ (a particular case of the unitary operator 
$v_{ra}$ introduced in [21,22]) through
          
	  \begin{eqnarray}
  v_{0a} |j , m \rangle = \left( 1 - \delta_{m,j} \right) q^{(j-m)a}
  |j , m+1 \rangle + \delta_{m,j} |j , -j \rangle  
          \nonumber 
	  \end{eqnarray} 
	    
or equivalently

\begin{eqnarray}
v_{0a} | k \rangle = q^{ka} | k-1 \rangle,
\nonumber 
\end{eqnarray}       

where $a = 0, 1, \ldots, 2j = d-1$, $q = \exp (2 \pi {i} / d)$, and 
$k-1$ should be understood modulo $d$ (i.e., $| -1 \rangle = | d -1 \rangle$).
It is evident that the operators $j^2$ and $v_{0a}$ commute so that the 
(complete) set $\{ j^2, v_{0a} \}$ constitutes an alternative to the set 
               $\{ j^2, j_z    \}$ [5--7,18]. The matrix $V_{0a}$ of the 
linear operator $v_{0a}$ on the basis $b_s \equiv B_d$ reads   
        
	\begin{eqnarray}
V_{0a} = 
\pmatrix{
0                    &    q^a &      0  & \ldots &       0 \cr
0                    &      0 & q^{2a}  & \ldots &       0 \cr
\vdots               & \vdots & \vdots  & \ldots &  \vdots \cr
0                    &      0 &      0  & \ldots & q^{2ja} \cr
1                    &      0 &      0  & \ldots &   0     \cr
},
        \nonumber 
	\end{eqnarray}
	
where the $d$ ($= 2j+1$) lines and $d$ columns are labeled in the order 
$|0 \rangle = |j , j \rangle, 
 |1 \rangle = |j , j-1 \rangle, \ldots, 
 |d - 1 \rangle = |j , -j \rangle$. 

By introducing the Hermitian operator $h$ via
        
	\begin{eqnarray}
   h |j , m \rangle = {\sqrt{ (j+m)(j-m+1) }} |j , m \rangle 
        \nonumber 
	\end{eqnarray}
	
or

        \begin{eqnarray}
   h |  k   \rangle = {\sqrt{ (d-1-k)(k+1) }} |  k   \rangle,  
        \nonumber 
	\end{eqnarray}
	
it can be shown that the three operators 
	  
	  \begin{eqnarray}
  j_+ = h           v_{0a},  \quad  
  j_- = v_{0a}^{\dagger} h,   \quad 
  j_z = \frac{1}{2} ( h^2 - v_{0a}^{\dagger} h^2 v_{0a} )
          \nonumber 
	  \end{eqnarray}
	  
satisfy the commutation relations
     
     \begin{eqnarray}
  \left[ j_z,j_{+} \right] = + j_{+},  \quad 
  \left[ j_z,j_{-} \right] = - j_{-},  \quad 
  \left[ j_+,j_- \right] = 2j_z. 
     \nonumber 
     \end{eqnarray}
     
Therefore, $j_+$, $j_-$ and $j_z$ span the Lie algebra su$_2$ of SU$_2$ 
and the operators $v_{0a}$ and $h$ realize a polar decomposition of su$_2$ 
[5--7,18--19,21--22].

We may ask what are the analogues of the vectors $|j , m \rangle$ 
in the $\{ j^2, v_{0a} \}$ scheme? In fact, the common eigenvectors 
of the commuting operators $v_{0a}$ and $j^2$ are [18--19,21--22] 
          
	  \begin{eqnarray}
|a \alpha \rangle = \frac{1}{\sqrt{2j+1}} \sum_{m = -j}^{j} 
q^{(j + m)(j - m + 1)a / 2 + (j + m)\alpha} | j , m \rangle
          \nonumber 
	  \end{eqnarray}
	   
or alternatively
          
	  \begin{eqnarray}
| a \alpha \rangle = \frac{1}{\sqrt{d}} \sum_{k = 0}^{d-1} 
q^{(d - k - 1)(k + 1)a / 2 - (k + 1) \alpha} | k \rangle, 
          \nonumber 
	  \end{eqnarray} 
	  
where $\alpha$ can take the values $\alpha = 0, 1, \ldots, 2j = d-1$. These vectors 
satisfy the eigenvalue equation 

\begin{eqnarray}
v_{0a} |a \alpha \rangle = q^{ja - \alpha} |a \alpha \rangle = 
                        q^{(d-1)a / 2 - \alpha} | a \alpha \rangle 
\nonumber 
\end{eqnarray}

that corresponds to a nondegenerate spectrum for the operator $v_{0a}$. For 
fixed $j$ and $a$ (with $2j \in {\bf N}$ and $a \in {\bf Z}_d$), the set 

\begin{eqnarray}
B_{0a} = \{ |a \alpha \rangle : \alpha = 0, 1, \ldots, d-1 \}
\nonumber 
\end{eqnarray} 

is an orthonormal basis for ${\cal E}(d)$, 
which is an alternative to the basis $B_d$.

\section{APPLICATION TO QUANTUM INFORMATION} 
We are now in a position to discuss some results of interest for
quantum information.

\subsection{Weyl pairs}

Besides the operator $v_{0a}$, it is interesting to define the unitary operator 
$z$ through 

        \begin{eqnarray}
 z = (v_{00})^{\dagger} v_{01}. 
	\nonumber 
	\end{eqnarray} 
	
This definition yields

          \begin{eqnarray}
z | j,m \rangle = q^{j-m} | j,m \rangle \Leftrightarrow  z | k \rangle = q^{k} | k \rangle.
          \nonumber 
	  \end{eqnarray}
	  
Therefore, we have the shift property

          \begin{eqnarray}
z | a \alpha \rangle = q^{-1} | a \alpha_1 \rangle, \quad \alpha_1 = \alpha -1. 
          \nonumber 
	  \end{eqnarray}
	  
Furthermore, we can show that $v_{0a}$ is connected to $z$ and $x = v_{00}$ by 

          \begin{eqnarray}
v_{0a} = x z^a.  
          \nonumber 
	  \end{eqnarray}
	  
The two isospectral operators $x$ (a shift operator when acting on 
$| j,m \rangle$ and a phase operator when acting on $| a \alpha \rangle$)
and $z$ (a phase operator when acting on $| j,m \rangle$ and a 
shift operator when acting on $| a \alpha \rangle$) are often 
called \emph{shift} operator and \emph{clock} operator, respectively, 
in quantum information and quantum computation. Note that for each 
of the operators $x$ and $z$, the \emph{shift} or \emph{clock} character 
depends on which state, $| j,m \rangle = |  k \rangle$ or 
$| a \alpha \rangle$, the operator acts. The operators $v_{0a}$ and $z$ 
satisfy 

\begin{eqnarray}
e^{-i \pi (d-1) a} \left( v_{0a} \right)^d = z^d = 1, \quad
v_{0a} z - q z v_{0a} = 0.
\nonumber 
\end{eqnarray}

In other words, the unitary operators $v_{0a}$ and $z$ are cyclic (up to a 
phase factor for $v_{0a}$ with $a \not= 0$) and do not commute. The pair 
$(x , z)$ corresponding to $a = 0$ is a Weyl pair in the sense that 
$x^d = z^d = 1$ and $x z = q z x$. Such a pair can be used as an integrity 
basis for constructing the Lie algebra of the unitary group U$_d$ [24].

In the basis $B_d$, the $d$-dimensional matrices $X$ and $Z$ of the 
linear operators $x$ and $z$ are given by 

        \begin{eqnarray}
X = 
\pmatrix{
0                    &      1 &  0      & \ldots &       0 \cr
0                    &      0 &  1      & \ldots &       0 \cr
\vdots               & \vdots &  \vdots & \ldots &  \vdots \cr
0                    &      0 &  0      & \ldots &       1 \cr
1                    &      0 &  0      & \ldots &       0 \cr
}
\nonumber 
\end{eqnarray}

and

\begin{eqnarray}
Z = 
\pmatrix{
1                    &      0 & 0      & \ldots &       0       \cr
0                    &      q & 0      & \ldots &       0       \cr
0                    &      0 & q^2    & \ldots &       0       \cr
\vdots               & \vdots & \vdots & \ldots &  \vdots       \cr
0                    &      0 & 0      & \ldots &       q^{d-1} \cr
}.
       \nonumber 
       \end{eqnarray}
       
For $d = 2 \Leftrightarrow j = 1/2$ ($\Rightarrow q = -1$), the 
matrices $V_{00}$, $V_{01}$, and $Z$ are connected to the Pauli 
matrices $\sigma_x$, $\sigma_y$, and $\sigma_z$ via 

\begin{eqnarray}
V_{00} = X  =     \sigma_x, \quad 
V_{01} = XZ = - i \sigma_y, \quad 
Z      = \sigma_z.
\nonumber 
\end{eqnarray}

For $d$ arbitrary, the pair ($X=V_{00}$, $Z=V_{00}^{\dagger} V_{01}$) are basic 
ingredients for generating generalized Pauli matrices and a generalized Pauli 
group $P_d$ of interest in quantum computation for quantum correcting codes 
[24,25]. 

\subsection{Mutually unbiased bases}

We now give two applications to MUBs.

First, in the general case where $d$ is arbitrary, we can 
check that 

          \begin{eqnarray}
| \langle k | a \alpha \rangle | = \frac{1}{\sqrt{d}}, \quad k, a, \alpha \in {\bf Z}_d
          \nonumber 
	  \end{eqnarray} 
	  
and 

          \begin{eqnarray}
| \langle 0 \alpha | 1 \beta \rangle | = \frac{1}{\sqrt{d}}, \quad \alpha, \beta \in {\bf Z}_d,
          \nonumber 
	  \end{eqnarray} 
	  
where we use $\langle \ | \ \rangle$ to denote the inner product 
in ${\cal E}(d)$. In the terminology of quantum information, the 
latter two equations mean that the bases $B_d$, $B_{00}$, and 
$B_{01}$ are three MUBs. This result is in agreement with the one 
according to which there exist at least three MUBs in arbitrary 
dimension, see for instance [17].
 
Second, in the special case where $d = p$ is a prime number, the 
latter equation can be extended to 

	  \begin{eqnarray}
| \langle a \alpha | b \beta \rangle | = 
\delta_{\alpha , \beta} \delta_{a , b} + \frac{1}{\sqrt{p}} (1 - \delta_{a , b}), 
\quad a, b, \alpha, \beta \in {\bf Z}_d.
          \nonumber 
	  \end{eqnarray} 
	  
The proof follows from the use of generalized quadratic Gauss 
sums [21]. Hence, the bases $B_{00}, B_{01}, \ldots, B_{0p-1}$, and 
$B_p$ constitute a (maximal) set of $p+1$ MUBs. This result is a 
particular case of the one according to which there exist a maximal 
set of $d+1$ MUBs when $d$ is the power of a prime number [8--17].

We continue with some typical examples of interest for quantum information and 
quantum computation.

\section{Examples} 

\subsection{The case $d=2$}
In this case, relevant for a spin $j = 1/2$ or for a qubit, we have $q = -1$ and 
$a, \alpha \in {\bf Z}_2$. The matrices of the operators $v_{0a}$ are 

     \begin{eqnarray}
V_{00} = 
\pmatrix{
  0     &1   \cr
  1     &0   \cr
} = \sigma_x, \quad 
V_{01} = 
\pmatrix{
  0     &-1  \cr
  1     &0   \cr
} = - i \sigma_y.
     \nonumber 
     \end{eqnarray} 
     
The $d+1=3$ MUBs $B_{2}$, $B_{00}$, and $B_{01}$ are the following.

\emph{The $B_{2}$ basis}:

           \begin{eqnarray}   
| 0 \rangle, \quad | 1 \rangle.
	   \nonumber 
	   \end{eqnarray}
	   
\emph{The $B_{00}$ basis}:	
   
	   \begin{eqnarray}
| 0 0 \rangle = \frac{1}{\sqrt{2}} \left(    | 0 \rangle + | 1 \rangle \right), \quad
| 0 1 \rangle = \frac{1}{\sqrt{2}} \left( -  | 0 \rangle + | 1 \rangle \right). 
	   \nonumber 
	   \end{eqnarray}
	   
\emph{The $B_{01}$ basis}:
	   
	   \begin{eqnarray}
| 1 0 \rangle = \frac{1}{\sqrt{2}} \left(  i | 0 \rangle + | 1 \rangle \right), \quad
| 1 1 \rangle = \frac{1}{\sqrt{2}} \left( -i | 0 \rangle + | 1 \rangle \right).
           \nonumber 
	   \end{eqnarray} 
	    
or, by using the spinorbitals $\alpha$ and $\beta$, we get 

\emph{The $B_{2}$ basis}:

\begin{eqnarray}
\alpha, \quad \beta. 
\nonumber 
\end{eqnarray}

\emph{The $B_{00}$ basis}:

\begin{eqnarray}
| 0 0 \rangle =    \frac{1}{\sqrt{2}} \left( \alpha +   \beta \right), \quad
| 0 1 \rangle = -  \frac{1}{\sqrt{2}} \left( \alpha -   \beta \right). 
\nonumber 
\end{eqnarray}

\emph{The $B_{01}$ basis}:

\begin{eqnarray}
| 1 0 \rangle =  i \frac{1}{\sqrt{2}} \left( \alpha - i \beta \right), \quad
| 1 1 \rangle = -i \frac{1}{\sqrt{2}} \left( \alpha + i \beta \right). 
\nonumber 
\end{eqnarray}

In terms of eigenvectors of the matrices $V_{0a}$, we must replace the 
state vectors $| a \alpha \rangle$ by column vectors. This leads to

\emph{The $B_{2}$ basis}:

\begin{eqnarray}
\alpha \to 
\pmatrix{
  1  \cr
  0  \cr
}, \quad
           \beta  \to 
\pmatrix{
  0  \cr
  1  \cr
}.
\nonumber 
\end{eqnarray}

\emph{The $B_{00}$ basis}:

\begin{eqnarray} 
| 0 0 \rangle \to    \frac{1}{\sqrt{2}} \pmatrix{
  1  \cr
  1  \cr
}, \quad
| 0 1 \rangle \to -  \frac{1}{\sqrt{2}} \pmatrix{
  1  \cr
 -1  \cr
}.
\nonumber 
\end{eqnarray}

\emph{The $B_{01}$ basis}:

\begin{eqnarray} 
| 1 0 \rangle \to  i \frac{1}{\sqrt{2}} \pmatrix{
  1  \cr
 -i  \cr
}, \quad
| 1 1 \rangle \to -i \frac{1}{\sqrt{2}} \pmatrix{
  1  \cr
  i  \cr
}. 
\nonumber 
\end{eqnarray}

\subsection{The case $d=3$}
This case corresponds to a spin $j=1$ or to a qutrit. Here, we have 
$q = \exp (2 \pi i / 3)$ and $a, \alpha \in {\bf Z}_3$. The 
matrices of the operators $v_{0a}$ are 

\begin{eqnarray}
V_{00} &=& 
\pmatrix{
  0     &1   &0 \cr
  0     &0   &1 \cr
  1     &0   &0 \cr
}  \nonumber \\ 
V_{01} &=&  
\pmatrix{
  0     &q   &0   \cr
  0     &0   &q^2 \cr
  1     &0   &0   \cr
}  \nonumber \\ 
V_{02} &=& 
\pmatrix{
  0     &q^2   &0 \cr
  0     &0     &q \cr
  1     &0     &0 \cr
}. \nonumber
\end{eqnarray}

The d+1 = 4 MUBs $B_{3}$, $B_{00}$, $B_{01}$, and $B_{02}$ are 
the following.

\emph{The $B_{3}$ basis}:

\begin{eqnarray}
| 0 \rangle,  \quad 
| 1 \rangle,  \quad  
| 2 \rangle.   
\nonumber 
\end{eqnarray}

\emph{The $B_{00}$ basis}:

\begin{eqnarray}
| 0 0 \rangle &=& \frac{1}{\sqrt{3}} \left(     | 0 \rangle +     | 1 \rangle + | 2 \rangle \right)  \nonumber \\
| 0 1 \rangle &=& \frac{1}{\sqrt{3}} \left( q^2 | 0 \rangle + q   | 1 \rangle + | 2 \rangle \right)  \nonumber \\ 
| 0 2 \rangle &=& \frac{1}{\sqrt{3}} \left( q   | 0 \rangle + q^2 | 1 \rangle + | 2 \rangle \right). \nonumber
\end{eqnarray}
\emph{The $B_{01}$ basis}:
\begin{eqnarray}
| 1 0 \rangle &=& \frac{1}{\sqrt{3}} \left( q   | 0 \rangle + q   | 1 \rangle + | 2 \rangle \right) \nonumber \\
| 1 1 \rangle &=& \frac{1}{\sqrt{3}} \left(     | 0 \rangle + q^2 | 1 \rangle + | 2 \rangle \right) \nonumber \\ 
| 1 2 \rangle &=& \frac{1}{\sqrt{3}} \left( q^2 | 0 \rangle +     | 1 \rangle + | 2 \rangle \right).\nonumber
\end{eqnarray}

\emph{The $B_{02}$ basis}:

\begin{eqnarray}
| 2 0 \rangle &=& \frac{1}{\sqrt{3}} \left( q^2 | 0 \rangle + q^2 | 1 \rangle + | 2 \rangle \right) \nonumber \\
| 2 1 \rangle &=& \frac{1}{\sqrt{3}} \left( q   | 0 \rangle +     | 1 \rangle + | 2 \rangle \right) \nonumber \\ 
| 2 2 \rangle &=& \frac{1}{\sqrt{3}} \left(     | 0 \rangle + q   | 1 \rangle + | 2 \rangle \right).\nonumber
\end{eqnarray}

This can be transcribed in terms of column vectors as follows. 

\emph{The $B_{3}$ basis}:

\begin{eqnarray}
| 0 \rangle \to 
\pmatrix{
  1  \cr
  0  \cr
  0  \cr
}, \quad
| 1 \rangle \to 
\pmatrix{
  0  \cr
  1  \cr
  0  \cr
}, \quad
| 2 \rangle \to 
\pmatrix{
  0  \cr
  0  \cr
  1  \cr
}.
\nonumber 
\end{eqnarray}

\emph{The $B_{00}$ basis}:

\begin{eqnarray}
| 0 0 \rangle \to   \frac{1}{\sqrt{3}} \pmatrix{
  1  \cr
  1  \cr
  1  \cr
}, \
| 0 1 \rangle \to   \frac{1}{\sqrt{3}} \pmatrix{
  q^2  \cr
  q    \cr
  1    \cr
}, \
| 0 2 \rangle \to   \frac{1}{\sqrt{3}} \pmatrix{
  q   \cr
  q^2 \cr
  1   \cr
}.
\nonumber 
\end{eqnarray}

\emph{The $B_{01}$ basis}:

\begin{eqnarray}
| 1 0 \rangle \to   \frac{1}{\sqrt{3}} \pmatrix{
  q  \cr
  q  \cr
  1  \cr
}, \
| 1 1 \rangle \to   \frac{1}{\sqrt{3}} \pmatrix{
  1    \cr
  q^2  \cr
  1    \cr
}, \
| 1 2 \rangle \to   \frac{1}{\sqrt{3}} \pmatrix{
  q^2    \cr
  1      \cr
  1      \cr
}.
\nonumber 
\end{eqnarray}

\emph{The $B_{02}$ basis}:

\begin{eqnarray}
| 2 0 \rangle \to   \frac{1}{\sqrt{3}} \pmatrix{
  q^2  \cr
  q^2  \cr
  1    \cr
}, \
| 2 1 \rangle \to   \frac{1}{\sqrt{3}} \pmatrix{
  q  \cr
  1  \cr
  1  \cr
}, \
| 2 2 \rangle \to   \frac{1}{\sqrt{3}} \pmatrix{
  1  \cr
  q  \cr
  1  \cr
}. 
\nonumber 
\end{eqnarray}

\subsection{The case $d=4$}
This case corresponds to a spin $j = 3/2$. Here, we have 
$q = i$ and $a, \alpha \in {\bf Z}_4$. The five bases 
$B_4$, $B_{00}$, $B_{01}$, $B_{02}$, and $B_{03}$ do not constitute 
a system of MUBs ($d=4$ is not a prime number). Nevertheless, it 
is possible to find $d+1 = 5$ MUBs 
because $d = 2^2$ is the power of a prime number. This can be 
achieved by replacing the space ${\cal E}(4)$ 
spanned by $\{ | 3/2 , m \rangle : m = 3/2, 1/2, -1/2, -3/2 \}$ 
by the tensor product space ${\cal E}(2) \otimes {\cal E}(2)$ 
spanned by the canonical (or computational) basis 

	\begin{eqnarray}
\{ \alpha \otimes \alpha, \alpha \otimes \beta, \beta \otimes \alpha, \beta \otimes \beta \}.
	\nonumber 
	\end{eqnarray}
	
The space ${\cal E}(2) \otimes {\cal E}(2)$ is associated with the coupling of two spin angular momenta 
$j_1 = 1/2$ and $j_2 = 1/2$ or two qubits (in the vector $u \otimes v$, $u$ and $v$ correspond to 
$j_1$       and $j_2$, respectively). An alternative basis for 
${\cal E}(2) \otimes {\cal E}(2)$ is the SU$_2$ adapted basis 

	\begin{eqnarray}
\{ \alpha \otimes \alpha, \frac{1}{2} (\alpha \otimes \beta + \beta \otimes \alpha), 
   \beta  \otimes \beta,  \frac{1}{2} (\alpha \otimes \beta - \beta \otimes
   \alpha) \},
	\nonumber 
	\end{eqnarray}
	
the vectors of which are well-known in the treatment 
of spin systems. Such a basis corresponds to the decomposition 

\begin{eqnarray}
(1/2) \otimes (1/2) = (1) \oplus (0)
\nonumber 
\end{eqnarray}

in terms of irreducible representation classes of SU$_2$. 
In the SU$_2$ adapted basis, the first three vectors are 
symmetric under the interchange $1 \leftrightarrow 2$ and 
describe a total angular momentum $J=1$ while the last 
one is antisymmetric and corresponds to $J=0$. It 
should be observed that the SU$_2$ adapted basis 
illustrates a connection between the special unitary 
group SU$_2$ and the permutation group S$_2$ (a particular 
case of the Schur-Weyl duality theorem between irreducible 
representation classes of the linear group GL$_d$ and the 
symmetric group S$_n$).  

In addition to the computational or canonical basis and the SU$_2$ 
adapted basis, it is possible to find other bases of 
${\cal E}(2) \otimes {\cal E}(2)$ which are mutually 
unbiased. As a matter of fact, $d=4$ MUBs, besides the 
computational basis 
$\{ \alpha \otimes \alpha, \alpha \otimes \beta, \beta \otimes \alpha, \beta \otimes \beta \}$, 
correspond to the eigenvectors 

	\begin{eqnarray}
|a b \alpha \beta \rangle = |a \alpha \rangle \otimes |b \beta \rangle 
	\nonumber 
	\end{eqnarray}
	
of the operators $w_{ab} = v_{0a} \otimes v_{0b}$ 
(the vectors $|a \alpha \rangle$ and $|b \beta \rangle$ refer 
to the two spaces ${\cal E}(2)$). As a result, we have 
the $d+1 = 5$ following MUBs where 
$\lambda = (1-i)/2$ and $\mu = (1+i)/2$.

\emph{The canonical basis}: 

	\begin{eqnarray}
\alpha \otimes \alpha, \quad \alpha \otimes \beta, \quad \beta \otimes \alpha, \quad \beta \otimes \beta
	\nonumber 
	\end{eqnarray} 
	
or in column vectors

	\begin{eqnarray}
\pmatrix{
1 \cr
0 \cr
0 \cr
0 \cr
}, \quad
\pmatrix{
0 \cr
1 \cr
0 \cr
0 \cr
}, \quad
\pmatrix{
0 \cr
0 \cr
1 \cr
0 \cr
}, \quad
\pmatrix{
0 \cr
0 \cr
0 \cr
1 \cr
}.
	\nonumber 
	\end{eqnarray}
	
\emph{The $w_{00}$ basis}:

\begin{eqnarray}
| 0 0 0 0 \rangle &=& \frac{1}{2} 
(\alpha \otimes \alpha + \alpha \otimes \beta + \beta \otimes \alpha + \beta \otimes \beta) \nonumber \\
| 0 0 0 1 \rangle &=& \frac{1}{2} 
(\alpha \otimes \alpha - \alpha \otimes \beta + \beta \otimes \alpha - \beta \otimes \beta) \nonumber \\
| 0 0 1 0 \rangle &=& \frac{1}{2} 
(\alpha \otimes \alpha + \alpha \otimes \beta - \beta \otimes \alpha - \beta \otimes \beta) \nonumber \\
| 0 0 1 1 \rangle &=& \frac{1}{2} 
(\alpha \otimes \alpha - \alpha \otimes \beta - \beta \otimes \alpha + \beta \otimes \beta) \nonumber      
\end{eqnarray}

or in column vectors

	\begin{eqnarray}
\frac{1}{2} \pmatrix{
1 \cr
1 \cr
1 \cr
1 \cr
}, \quad
\frac{1}{2} \pmatrix{
1 \cr
-1 \cr
1 \cr
-1 \cr
}, \quad
\frac{1}{2} \pmatrix{
1 \cr
1 \cr
-1 \cr
-1 \cr
}, \quad
\frac{1}{2} \pmatrix{
1 \cr
-1 \cr
-1 \cr
1 \cr
}.
	\nonumber 
	\end{eqnarray}
	
\emph{The $w_{11}$ basis}:

\begin{eqnarray}
| 1 1 0 0 \rangle &=& \frac{1}{2} 
(\alpha \otimes \alpha + i \alpha \otimes \beta + i \beta \otimes \alpha - \beta \otimes \beta) \nonumber  \\
| 1 1 0 1 \rangle &=& \frac{1}{2} 
(\alpha \otimes \alpha - i \alpha \otimes \beta + i \beta \otimes \alpha + \beta \otimes \beta) \nonumber  \\
| 1 1 1 0 \rangle &=& \frac{1}{2} 
(\alpha \otimes \alpha + i \alpha \otimes \beta - i \beta \otimes \alpha + \beta \otimes \beta) \nonumber  \\
| 1 1 1 1 \rangle &=& \frac{1}{2} 
(\alpha \otimes \alpha - i \alpha \otimes \beta - i \beta \otimes \alpha - \beta \otimes \beta) \nonumber  
\end{eqnarray}

or in column vectors

	\begin{eqnarray}
\frac{1}{2} \pmatrix{
1 \cr
i \cr
i \cr
-1 \cr
}, \quad
\frac{1}{2} \pmatrix{
1 \cr
-i \cr
i \cr
1 \cr
}, \quad
\frac{1}{2} \pmatrix{
1 \cr
i \cr
-i \cr
1 \cr
}, \quad
\frac{1}{2} \pmatrix{
1 \cr
-i \cr
-i \cr
-1 \cr
}.
	\nonumber 
	\end{eqnarray}
	
\emph{The $w_{01}$ basis}:

\begin{eqnarray}
\lambda | 0 1 0 0 \rangle + \mu | 0 1 1 1 \rangle &=& \frac{1}{2} 
(\alpha \otimes \alpha + \alpha \otimes \beta - i \beta \otimes \alpha + i \beta \otimes \beta) \nonumber  \\
\mu | 0 1 0 0 \rangle + \lambda | 0 1 1 1 \rangle &=& \frac{1}{2} 
(\alpha \otimes \alpha - \alpha \otimes \beta + i \beta \otimes \alpha + i \beta \otimes \beta) \nonumber  \\
\lambda | 0 1 0 1 \rangle + \mu | 0 1 1 0 \rangle &=& \frac{1}{2} 
(\alpha \otimes \alpha - \alpha \otimes \beta - i \beta \otimes \alpha - i \beta \otimes \beta) \nonumber  \\
\mu | 0 1 0 1 \rangle + \lambda | 0 1 1 0 \rangle &=& \frac{1}{2} 
(\alpha \otimes \alpha + \alpha \otimes \beta + i \beta \otimes \alpha - i \beta \otimes \beta) \nonumber      
\end{eqnarray}

or in column vectors

	\begin{eqnarray}
\frac{1}{2} \pmatrix{
1 \cr
1 \cr
-i \cr
i \cr
}, \quad
\frac{1}{2} \pmatrix{
1 \cr
-1 \cr
i \cr
i \cr
}, \quad
\frac{1}{2} \pmatrix{
1 \cr
-1 \cr
-i \cr
-i \cr
}, \quad
\frac{1}{2} \pmatrix{
1 \cr
1 \cr
i \cr
-i \cr
}.
	\nonumber 
	\end{eqnarray}
	
\emph{The $w_{10}$ basis}:

\begin{eqnarray}
\lambda | 1 0 0 0 \rangle + \mu | 1 0 1 1 \rangle &=& \frac{1}{2} 
(\alpha \otimes \alpha - i \alpha \otimes \beta + \beta \otimes \alpha + i \beta \otimes \beta) \nonumber \\
\mu | 1 0 0 0 \rangle + \lambda | 1 0 1 1 \rangle &=& \frac{1}{2} 
(\alpha \otimes \alpha + i \alpha \otimes \beta - \beta \otimes \alpha + i \beta \otimes \beta) \nonumber \\
\lambda | 1 0 0 1 \rangle + \mu | 1 0 1 0 \rangle &=& \frac{1}{2} 
(\alpha \otimes \alpha + i \alpha \otimes \beta + \beta \otimes \alpha - i \beta \otimes \beta) \nonumber \\
\mu | 1 0 0 1 \rangle + \lambda | 1 0 1 0 \rangle &=& \frac{1}{2} 
(\alpha \otimes \alpha - i \alpha \otimes \beta - \beta \otimes \alpha - i \beta \otimes \beta) \nonumber 
\end{eqnarray}

or in column vectors

	\begin{eqnarray}
\frac{1}{2} \pmatrix{
1 \cr
-i \cr
1 \cr
i \cr
}, \quad
\frac{1}{2} \pmatrix{
1 \cr
i \cr
-1 \cr
i \cr
}, \quad
\frac{1}{2} \pmatrix{
1 \cr
i \cr
1 \cr
-i \cr
}, \quad
\frac{1}{2} \pmatrix{
1 \cr
-i \cr
-1 \cr
-i \cr
}.
	\nonumber 
	\end{eqnarray}
	
It is to be noted that the vectors of the $w_{00}$ and $w_{11}$ bases are not intricated 
(i.e., each vector is the direct product of two vectors) while the vectors of the $w_{01}$ 
and $w_{10}$ bases are intricated (i.e., each vector is not the direct product of two vectors). 

To be more precise, the degree of intrication of the state 
vectors for the bases $w_{00}$, $w_{11}$, $w_{01}$, and $w_{10}$ 
can be determined in the following way. In arbitrary dimension $d$, 
let 

\begin{eqnarray}
| \Phi \rangle = \sum_{k = 0}^{d-1} \sum_{l = 0}^{d-1} a_{kl} | k \rangle \otimes | l \rangle
\nonumber 
\end{eqnarray}

be a double qudit state vector. Then, it can be shown that the 
determinant of the $d \times d$ matrix $A = (a_{kl})$ satisfies 

\begin{eqnarray}
0 \leq |\det A| \leq \frac{1}{\sqrt{d^d}}
\nonumber 
\end{eqnarray}

as proved in the Albouy thesis [26]. The case $\det A = 0$ 
corresponds to the absence of \emph{global} intrication while the case 

\begin{eqnarray}
|\det A| = \frac{1}{\sqrt{d^d}}
\nonumber 
\end{eqnarray} 

corresponds to a maximal 
intrication. As an illustration,  we   obtain   that all the state vectors 
for $w_{00}$ and $w_{11}$ are not intricated and that all the state vectors 
for $w_{01}$ and $w_{10}$ are maximally intricated.

\section{CONCLUSION}

We presented some results concerning bases of interest 
for  spin  systems and quantum information.  It should
be noted that the bases derived in the present work are adapted to 
cyclic symmetry and can be used for describing finite or infinite cyclic 
systems. As an extension of this paper and of its companion paper [24], it would 
be interesting to examine when $d$ is a composite number how the Weyl 
pair $(X , Z)$ and the corresponding generalized Pauli group $P_d$ are 
connected to the Weyl pairs and the corresponding generalized Pauli group 
for each of the composite dimension. 

We close with a comment on the group $P_d$. Such 
a group can be generated from the Weyl pair 
$(X,Z)$. Indeed, the group $P_d$ is a finite group of order $d^3$ 
with elements $q^aX^bZ^c$ (where $a,b,c \in {\bf Z}_d$) and matrix multiplication 
for group law. It is a subgroup of the unitary group U$_d$. The normaliser of 
$P_d$ in U$_d$ is a Clifford-type group in $d$ dimensions noted $C_d$. 
More precisely, $C_d$ is the set 
$\{ U \in {\rm U}_d | U P_d U^{\dagger} = P_d \}$ endowed with matrix 
multiplication. The Pauli group $P_d$ as well as any other invariant subgroup 
of $C_d$ can be used for stabilizing errors in quantum computing. These
concepts are very important in the case of n-qubit systems (corresponding 
to $d = 2^n$). These matters, in connection with the decomposition of 
$P_{2^n}$ in terms of $P_{2}$, are the object of a future work. 

\begin{ack}
The present paper is based on an invited talk 
to the International Workshop on New Trends in Science 
and Technology (3--4 November 2008, Ankara, Turkey). The 
author thanks the organizers for making possible this 
interesting pluri-disciplinary workshop. 
\end{ack}

\bibliography{ifacconf}

\end{document}